\documentclass[12pt]{iopart}

\usepackage{graphicx}
\usepackage{multirow}
\usepackage{dcolumn}
\usepackage{cite}

\begin{document}

\title{Critical parameters for non-hermitian Hamiltonians}
\author{Francisco M Fern\'andez \ and Javier Garcia}

\address{INIFTA (UNLP, CCT La Plata-CONICET), Divisi\'on Qu\'imica Te\'orica,
Blvd. 113 S/N,  Sucursal 4, Casilla de Correo 16, 1900 La Plata,
Argentina}

\ead{fernande@quimica.unlp.edu.ar}

\maketitle

\begin{abstract}
We calculate accurate critical parameters for a class of non-hermitian
Hamiltonians by means of the diagonalization method. We study three
one-dimensional models and two perturbed rigid rotors with PT symmetry. One
of the latter models illustrates the necessity of a more general condition
for the appearance of real eigenvalues that we also discuss here.
\end{abstract}

\section{Introduction}

\label{sec:intro}

There has recently been interest in PT-symmetric Hamiltonians that
exhibit real eigenvalues for a range of values of a potential
parameter. Some of them are anharmonic oscillators\cite
{DP98,FGRZ98,DT00,BBMSS01,H01,HKWT01,HW03,BW12,HV13} as well as
models with Dirichlet\cite{RSM07,Z01,ZL01} periodic and
anti-periodic boundary conditions\cite{JZ04,BK11}.

Among the methods used for the study of such models we mention the
WKB approximation\cite{DT00,BBMSS01}, the eigenvalue moment
method\cite{H01,HW03}, the multiscale reference function
analysis\cite{HKWT01}, the diagonalization method (DM)\cite{BW12}
and the orthogonal polynomial projection quantization (OPPQ) (an
improved Hill-determinant method)\cite {HV13}.

For some particular values of the potential parameter the spectrum of those
PT-symmetric Hamiltonians exhibits critical points where two real
eigenvalues coalesce and emerge as complex conjugate eigenvalues. Such
critical points are also known as exceptional points\cite{HS90,H00,HH01,H04}.

The purpose of this paper is the analysis of the critical points for a
variety of simple models. The calculation is based on a well known simple
and quite efficient application of the DM\cite{HS90}. In section~\ref
{sec:pseudo_sym} we propose a somewhat more general condition for the
existence of real eigenvalues (unbroken symmetry)\cite{BMW03,BBRR04} that is
suitable for models with degenerate states. In section~\ref{sec:examples} we
present three one-dimensional examples already discussed earlier by other
authors. In section~\ref{sec:diagonalization} we outline the procedure for
the calculation of critical points based on the DM. In section~\ref
{sec:pert_theor} we apply perturbation theory to one of the models and
discuss the convergence of the perturbation series for the eigenvalues by
comparison with the accurate results produced by the DM. In section~\ref
{sec:planar_rotor} we discuss a PT-symmetric perturbed planar rigid rotor
that was studied earlier as an example with E2 algebra\cite{BK11}. In
section~\ref{sec:3D_rotor} we discuss a non-hermitian perturbed
three-dimensional rigid rotor that was not treated before as far as we know.
This most interesting model illustrates the generalized condition for real
eigenvalues mentioned above. Finally, in section~\ref{sec:conclusions} we
summarize the main results and draw conclusions.

\section{PT Symmetry}

\label{sec:pseudo_sym}

It is well known that a wide class of non-hermitian Hamiltonians with
unbroken PT symmetry exhibit real spectra\cite{BMW03,BBRR04}. In general,
they are invariant under an antilinear or antiunitary transformation of the
form $\hat{A}^{-1}\hat{H}\hat{A}=\hat{H}$. The antiunitary operator $\hat{A}$
satisfies\cite{W60}
\begin{eqnarray}
\hat{A}\left( \left| f\right\rangle +\left| g\right\rangle \right) &=&\hat{A}%
\left| f\right\rangle +\hat{A}\left| g\right\rangle  \nonumber \\
\hat{A}c\left| f\right\rangle &=&c^{*}\hat{A}\left| f\right\rangle ,
\label{eq:antiunitary_1}
\end{eqnarray}
for any pair of vectors $\left| f\right\rangle $ and $\left| g\right\rangle $
and arbitrary complex number $c$, where the asterisk denotes complex
conjugation. This definition is equivalent to
\begin{equation}
\left\langle \hat{A}f\right. \left| \hat{A}g\right\rangle =\left\langle
f\right. \left| g\right\rangle ^{*}  \label{eq:antiuniary_2}
\end{equation}

It follows from the antiunitary invariance mentioned above that $[\hat{H},%
\hat{A}]=0$. Therefore, if $\left| \psi \right\rangle $ is an eigenvector of
$\hat{H}$ with eigenvalue $E$
\begin{equation}
\hat{H}\left| \psi \right\rangle =E\left| \psi \right\rangle ,
\end{equation}
we have
\begin{equation}
\lbrack \hat{H},\hat{A}]\left| \psi \right\rangle =\hat{H}\hat{A}\left| \psi
\right\rangle -\hat{A}\hat{H}\left| \psi \right\rangle =\hat{H}\hat{A}\left|
\psi \right\rangle -E^{*}\hat{A}\left| \psi \right\rangle =0.
\end{equation}
This equation merely tell us that if $\left| \psi \right\rangle$
is eigenvector of $\hat{H}$ with eigenvalue $E$ then
$\hat{A}\hat{H}\left| \psi \right\rangle$ is eigenvector with
eigenvalue $E^{*}$. Consequently, $E$ is real if
\begin{equation}
\hat{H}\hat{A}\left| \psi \right\rangle =E\hat{A}\left| \psi \right\rangle ,
\label{eq:gen_unb_sym}
\end{equation}
that contains the condition of unbroken symmetry required by Bender et al%
\cite{BMW03,BBRR04}
\begin{equation}
\hat{A}\left| \psi \right\rangle =\lambda \left| \psi \right\rangle
\label{eq:unb_sym}
\end{equation}
as a particular case. Note that equation (\ref{eq:gen_unb_sym}) applies to
the case in which $\hat{A}\left| \psi \right\rangle $ is a linear
combination of degenerate eigenvectors of $\hat{H}$ with eigenvalue $E$.

If $\hat{K}$ is an antilinear operator such that
$\hat{K}^{2}=\hat{1}$ (for example, the complex conjugation
operator) then it follows from (\ref {eq:antiuniary_2}) that
$\hat{A}\hat{K}=\hat{U}$ is unitary ($\hat{U}^{\dagger
}=\hat{U}^{-1}$). In other words, any antilinear operator
$\hat{A}$ can be written as a product of a unitary operator and
the complex conjugation operation\cite{W60}. In most of the
non-hermitian models studied $\hat{U}^{-1}=\hat{U}$ that results
in $\hat{A}^{2}=\hat{1}$ (as in the case of the parity operator
$\hat{U}=\hat{P}$ that gives rise to PT symmetry)\cite
{BMW03,BBRR04}.

\section{Some simple one-dimensional examples}

\label{sec:examples}

In this section we consider three examples of the Schr\"{o}dinger equation
\begin{eqnarray}
\hat{H}\psi &=&E\psi  \nonumber \\
\hat{H} &=&\hat{p}^{2}+\hat{V}(x),  \label{eq:Schrodinger}
\end{eqnarray}
with eigenvalues $E_{0}<E_{1}<\ldots $.

The first one\cite{DT00,H01,HKWT01}
\begin{equation}
\hat{H}=\hat{p}^{2}+i\hat{x}^{3}+ia\hat{x},  \label{eq:Hcubic}
\end{equation}
exhibits an infinite set of critical values $0>a_{0}>a_{1}>...>a_{n}>...$ of
$a$ so that $E_{2n}=E_{2n+1}$ at $a=a_{n}$. Both eigenvalues are real when $%
a>a_{n}$ and become complex conjugate numbers when $a<a_{n}$. The
eigenfunctions $\psi _{2n}$ and $\psi _{2n+1}$ are linearly dependent at the
exceptional point $a=a_{n}$\cite{HS90,H00,HH01,H04}.

The second example is\cite{DP98,BBMSS01,HW03}
\begin{equation}
\hat{H}=\hat{p}^{2}+\hat{x}^{4}+ia\hat{x}.  \label{eq:Hcuartic}
\end{equation}
If $\hat{P}$ denotes the parity operator we have $\hat{P}\hat{H}(a)\hat{P}=%
\hat{H}(-a)$ so that $E(-a)=E(a)$. Because of this property of the
eigenvalues the crossings $E_{2n}=E_{2n+1}$ take place at $\pm a_{n}$, where
$0<a_{0}<a_{1}<\ldots <a_{n}<...$. In this case the pair of coalescing
eigenvalues become complex conjugate numbers when $|a|>a_{n}$

The third example is given by
\begin{equation}
\hat{H}=\hat{p}^{2}+ia\hat{x},  \label{eq:Hbox}
\end{equation}
with the boundary conditions $\psi (\pm 1)=0$. In this case we also find
that the crossings take place at $\pm a_{n}$, $a_{n}>0$ as in the preceding
one. Because of physical reasons Rubinstein et al\cite{RSM07} considered
only the half line $a>0$.

\section{Diagonalization method}

\label{sec:diagonalization}

In order to solve the Schr\"{o}dinger equation (\ref{eq:Schrodinger}) we
resort to a matrix representation of the Hamiltonian operator $%
H_{ij}=\left\langle i\right| \hat{H}\left| j\right\rangle $ in an
appropriate orthonormal basis set $\{\left| j\right\rangle,\,j=0,1,\ldots \}$%
. We obtain the eigenvalues from the roots of the characteristic polynomial
given by the secular determinant $D(E,a)=\left| \mathbf{H}-E\mathbf{I}%
\right| =0$, where $\mathbf{H}$ is an $N\times N$ matrix with elements $%
H_{ij}$, $i,j=0,1,\ldots ,N-1$ and $\mathbf{I}$ is the $N\times N$ identity
matrix. We look for those roots of the characteristic polynomial that
converge as $N$ increases. The characteristic polynomial gives us either $%
E(a)$ or $a(E)$.

In all the examples discussed here the critical parameters are given by $%
a_{n}=a(e_{n})$, where
\begin{equation}
\left. \frac{da}{dE}\right| _{E=e_{n}}=0,  \label{eq:crit_cond}
\end{equation}
and $E_{2n}(a_{n})=E_{2n+1}(a_{n})=e_{n}$. Therefore, we can obtain the
critical parameters approximately from the set of polynomial equations $%
\{D(E,a)=0,\partial D(E,a)/\partial E=0\}$\cite{HS90}. We look for pairs of
roots $(a_{n,N},e_{n,N})$ that converge as $N\rightarrow \infty $.

The eigenvectors of the harmonic oscillator $\hat{H}_{0}=\hat{p}^{2}+\hat{x}%
^{2}$ are a suitable basis set for the first two examples (\ref{eq:Hcubic})
and (\ref{eq:Hcuartic}), and for the third one (\ref{eq:Hbox}) we choose
\begin{equation}
\phi _{n}(x)=\left\langle x\right| \left. n\right\rangle =\sin \left( \frac{%
n\pi (x+1)}{2}\right) ,\,n=1,2,\ldots  \label{eq:phi_box}
\end{equation}

Before proceeding with the discussion of the examples we want to stress that
the DM is a simple and most efficient approach for the accurate calculation
of the eigenvalues and eigenfunctions of those PT-symmetric oscillators with
eigenfunctions that vanish exponentially along the real $x$ axis. In order
to illustrate this point we compare the DM with the recently developed OPPQ%
\cite{HV13}. As an example we choose the PT-symmetric oscillator $\hat{H}=%
\hat{p}^{2}+i\hat{x}^{3}$ because Handy and Vrinceanu\cite{HV13} showed OPPQ
results of increasing order of accuracy for this model. Although both
methods resort to the same Gaussian function and Hermite polynomials, Table~%
\ref{tab:ix3_conv} shows that the rate of convergence of the DM is
noticeably greater. It is striking that the DM of order $N$ appears to be
nearly as accurate as the OPPQ of order $N+20$.

Tables \ref{tab:CP_cubic}, \ref{tab:CP_cuartic} and
~\ref{tab:CP_box} show the first critical parameters for the
examples (\ref{eq:Hcubic}), (\ref {eq:Hcuartic}) and
(\ref{eq:Hbox}) calculated with $N\leq300$, $N\leq300$ and $N=100$
basis functions, respectively. With those results we carried out
nonlinear regressions of the form
\begin{equation}
a_{n}=b+ce_{n}^{s},
\end{equation}
and obtained
\begin{eqnarray}
b &=&-0.324\pm 0.015  \nonumber \\
c &=&-1.9288\pm 0.0083  \nonumber \\
s &=&0.6751\pm 0.0011,  \label{eq:nonreg_cubic}
\end{eqnarray}
for (\ref{eq:Hcubic}),
\begin{eqnarray}
b &=&0.407\pm 0.010  \nonumber \\
c &=&1.1540\pm 0.0048  \nonumber \\
s &=&0.7555\pm 0.0010,  \label{eq:nonreg_cuartic}
\end{eqnarray}
for (\ref{eq:Hcuartic}) and

\begin{eqnarray}
b &=&-0.00028\pm 0.00016  \nonumber \\
c &=&1.732092\pm 9.5\times 10^{-6}  \nonumber \\
s &=&0.9999951\pm 9.4\times 10^{-7},  \label{eq:nonreg_box}
\end{eqnarray}
for (\ref{eq:Hbox}). The parameters in equation (\ref{eq:nonreg_cuartic})
are in good agreement with the WKB ones\cite{BBMSS01} which suggests that
even the first critical parameters for that model exhibit the large-$e_{n}$
asymptotic behaviour given exactly by the WKB method. The nonlinear
regression appears to be most accurate for the example (\ref{eq:Hbox}) where
it seems that $a_{n}=1.7321e_{n}$. It seems that $e_{n}$ is always
approximately between $E_{2n-1}(a=0)$ and $E_{2n}(a=0)$ and, therefore,
increases asymptotically as $n^{2}$. Consequently, $a_{n}$ behaves
approximately in the same way.

\begin{figure}[tbp]
\begin{center}
\includegraphics[width=9cm]{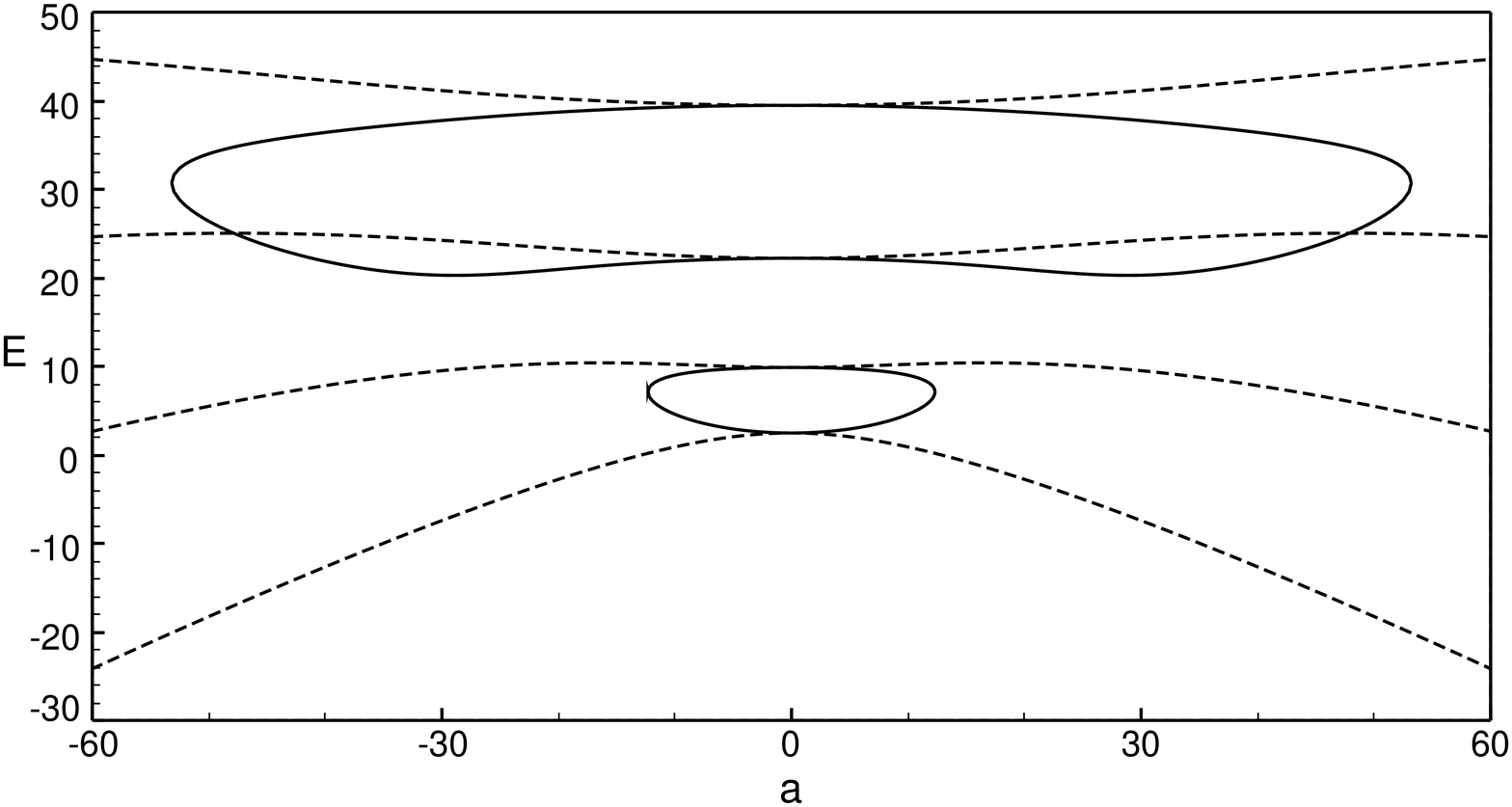}
\end{center}
\caption{First four eigenvalues $E_n(g)$ for the model (\ref{eq:Hbox}) with $%
g=a$ (dashed line) and $g=ia$ (solid line) }
\label{fig:PTB}
\end{figure}

In the discussion below we sometimes find it convenient to write $g$ for $ia$
and consider $g$ complex. Figure~\ref{fig:PTB} shows $E_{n}(g)$, $n=1,2,3,4$
for the example (\ref{eq:Hbox}) for $g$ real and purely imaginary. We will
discuss this case with more detail in the next section.

\section{Perturbation theory}

\label{sec:pert_theor}

Delabaere and Trinh\cite{DT00} derived the exact
Rayleigh-Schr\"{o}dinger series asymptotic to the eigenvalues of
the Hamiltonian (\ref{eq:Hcubic}) for large $a$. For the three
examples discussed in section~\ref{sec:examples} it is also
possible to obtain a perturbation series for small $a$ (see, for
example, Fern\'andez et al\cite{FGRZ98}). In all of them the
Taylor series for $E_{n}$ about $a=0$ exhibits a finite nonzero
radius of convergence (see, for example, page 111 in
reference\cite{F01} and references therein). The three
Hamiltonians are PT symmetric when $g$ is imaginary and
(\ref{eq:Hcuartic}) and (\ref{eq:Hbox}) are Hermitian when $g$ is
real. The perturbation series for the eigenvalues of either (\ref
{eq:Hcuartic}) or (\ref{eq:Hbox}) reads
\begin{equation}
E_{n}(g)=\sum_{j=0}^{\infty }E_{n,j}g^{2j}.  \label{eq:En_g2_series}
\end{equation}
We can calculate the coefficients $E_{n,j}$ approximately for the former and
exactly for the latter. By means of a variety of well known methods\cite{F01}
we easily obtain
\begin{eqnarray}
E_{n} &=&\frac{1}{2}b_{n}+\,{\frac{\left( 2\,b_{n}-15\right) {g}^{2}}{%
12b_{n}^{2}}}+\,{\frac{\left( b_{n}^{2}-105\,b_{n}+495\right) {g}^{4}}{%
18b_{n}^{5}}}  \nonumber \\
&&+{\frac{\left( 2\,b_{n}^{3}-825b_{n}^{2}+23400\,b_{n}-95625\right) {g}^{6}%
}{36b_{n}^{8}}+\ldots ,}  \label{eq:RS_ser_box}
\end{eqnarray}
where $b_{n}=n^{2}\pi ^{2}/2$. The radius of convergence of the perturbation
series for both $E_{2n-1}$ and $E_{2n}$, $n=1,2,\ldots $ cannot be greater
than $a_{n}$ because the two eigenvalues coalesce at the exceptional branch
points $g=\pm ia_{n}$.

\begin{figure}[tbp]
\begin{center}
\par
\includegraphics[width=9cm]{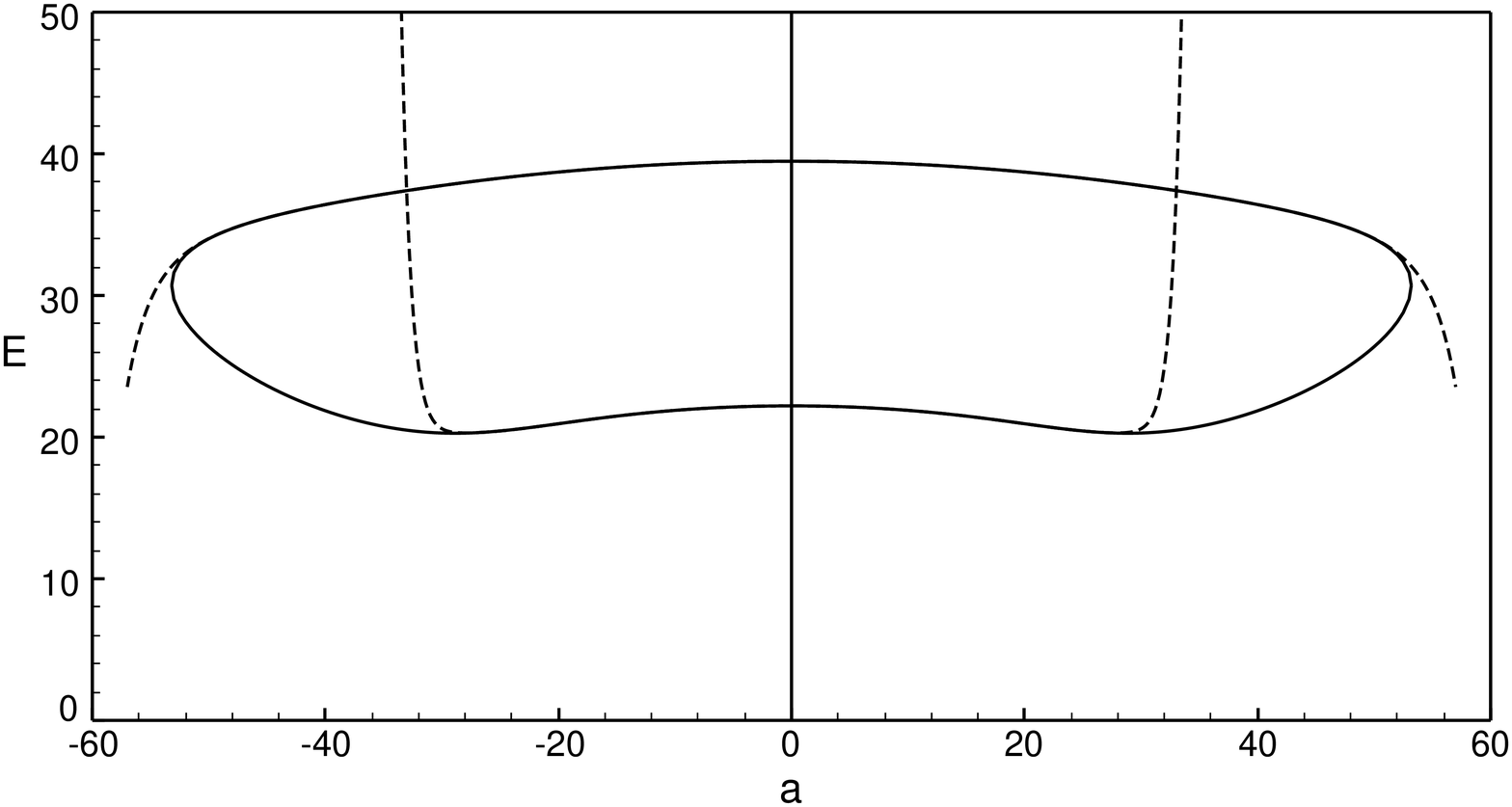}
\par
\includegraphics[width=9cm]{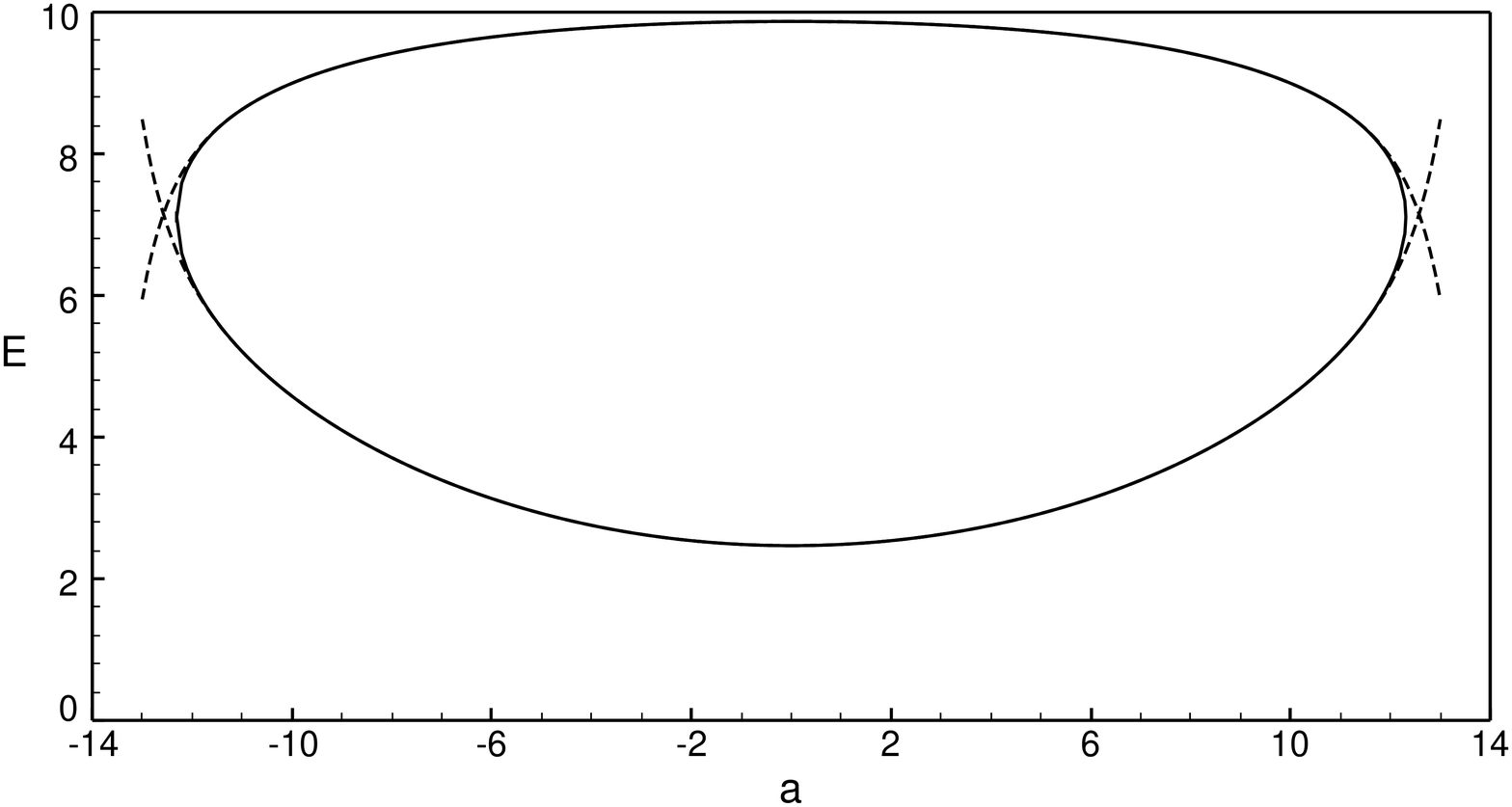}
\par
\end{center}
\caption{First four eigenvalues $E_n(g)$ for the model (\ref{eq:Hbox}) with $%
g=ia$ calculated by the diagonalization method (solid line) and perturbation
theory (dashed line) }
\label{fig:PTBPT}
\end{figure}

Figure~\ref{fig:PTBPT} shows the first four eigenvalues of the
problem (\ref {eq:Hbox}) when $g=ia$ calculated by means of the DM
and by perturbation theory of order 20. We appreciate that there
is a good agreement between both approaches for the first two
eigenvalues for almost all the values of $-a_{1}<a<a_{1}$ except
close to the crossings where perturbation theory is expected to
fail. The situation appears to be quite similar for the fourth
eigenvalue but the behaviour of the perturbation series for the
third eigenvalue strongly suggests that its radius of convergence
may be considerably smaller than $a_{2}$.

If $g=\pm ia_{n}$ were the singularities closest to the origin, one could
obtain them from the perturbation coefficients $E_{n,j}$ as follows:\cite
{F01}
\begin{equation}
a_{n}=\lim\limits_{k\rightarrow \infty }\left| \frac{\left( 1/2-k\right)
E_{2n-1,k}}{(k+1)E_{2n-1,k+1}}\right| ^{1/2}=\lim\limits_{k\rightarrow
\infty }\left| \frac{\left( 1/2-k\right) E_{2n,k}}{(k+1)E_{2n,k+1}}\right|
^{1/2}.  \label{eq:an_PT}
\end{equation}
Table~\ref{tab:PT_Crit} shows that $a_{1}(k)=$ $\left|
\frac{\left( 1/2-k\right) E_{1,k}}{(k+1)E_{1,k+1}}\right| ^{1/2}$
already converges towards the result in Table~\ref{tab:CP_box} as
$k$ increases, and we obtain identical results with the
coefficient $E_{2,k}$ as expected. However, the sequences with
$E_{n,k}$ do not converge when $n>2$ which suggests that there may
be other branch points on the complex $g$-plane closest to the
origin. For example, $E_{3}(g)$ exhibits branch points at
$g_{c}=\pm 11.48088661+26.24188126i$ and also at $g_{c}^{*}$ that
are closer to the origin than $g_{2}=ia_{2}$
($|g_{c}|=28.64344759<a_{2}$). As already mentioned above,
equation (\ref{eq:an_PT}) is only suitable for a branch point on
the imaginary axis\cite{F01} and therefore does not converge in
the latter case. The branch points at $g_{c}$ and $g_{c}^{*}$
account for the behaviour of the perturbation series for
$E_{3}(g)$ in Fig.~\ref {fig:PTBPT} discussed above.

\section{Non-hermitian perturbed planar rigid rotor}

\label{sec:planar_rotor}

In this section we consider a simple model with periodic boundary conditions
that we prefer to treat separately from those in section~\ref{sec:examples}.

Bender and Kalvecks\cite{BK11} studied the eigenvalues of
\begin{equation}
-\psi ^{\prime \prime }(\theta )+g\cos (\theta )\psi (\theta )=E\psi (\theta
),  \label{eq:rotor_2D_eq}
\end{equation}
with periodic $\psi (\theta +2\pi )=\psi (\theta )$ and anti-periodic $\psi
(\theta +2\pi )=-\psi (\theta )$ boundary conditions. This equation is a
particular case of\cite{BK11}
\begin{eqnarray}
\hat{H}\psi &=&E\psi ,  \nonumber \\
\hat{H} &=&\hat{J}^{2}+V(g,\theta ),\;\hat{J}=-i\frac{d}{d\theta },
\label{eq:rotor_2D_op}
\end{eqnarray}
when $V(g,\theta )=g\cos (\theta )$.

By means of the unitary operator $\hat{U}$ that produces the transformation $%
\hat{U}^{\dagger }\theta $ $\hat{U}=\theta +\pi $, $\hat{U}^{\dagger }\hat{J}
$ $\hat{U}=\hat{J}$ we can construct the antiunitary operator $\hat{A}=\hat{U%
}\hat{T}=\hat{T}\hat{U}$, where $\hat{T}$ is the time-inversion operator, as
indicated in section~\ref{sec:pseudo_sym}. Since $A^{-1}\hat{H}\hat{A}=\hat{H%
}$ when $g=ia$ is purely imaginary we expect real eigenvalues for some real
values of $a$.

Here we consider only periodic boundary conditions and transform equation (%
\ref{eq:rotor_2D_eq}) into the Mathieu equation\cite{AS72} by means of the
transformations $\theta =2x$, $E^{BK}=E/4$ and $g^{BK}=g/2$, so that
\begin{equation}
\varphi ^{\prime \prime }(x)+\left[ E-2g\cos (2x)\right] \varphi (x)=0,
\label{eq:Mathieu}
\end{equation}
where $\varphi (x)=\psi (2x)$. The even and odd solutions to this equation
can be expanded in the Fourier series
\begin{eqnarray}
\varphi _{e}(x) &=&\sum_{m=0}^{\infty }A_{2m}\cos (2mx),  \nonumber \\
\varphi _{o}(x) &=&\sum_{m=1}^{\infty }B_{2m}\sin (2mx),
\label{eq:Mathieu_series}
\end{eqnarray}
respectively, where the coefficients $A_{2m}$ and $B_{2m}$ can be calculated
by means of simple three-term recurrence relations\cite{AS72}. We can
efficiently calculate accurate eigenvalues from either the secular
determinant, as discussed in section~\ref{sec:diagonalization}, or the
truncation conditions $A_{2N}=0$ and $B_{2N}=0$ for sufficiently large
values of $N$. We denote $E_{e,n}$ $n=0,1,\ldots $ and $E_{o,n}$, $%
n=1,2,\ldots $ the eigenvalues of the even and odd solutions, respectively.
Obviously, $E_{e,n}=E_{o,n}=4n^{2}$, $n=1,2,\ldots $, when $g=0$.

The results of Bender and Kalveks\cite{BK11} suggest that pairs of
eigenvalues $(E_{e,2n},E_{e,2n+1})$ and $(E_{o,2n+1},E_{o,2n})$, $%
n=0,1,\ldots $ coalesce at $\pm a_{e,n}$ and $\pm a_{o,n}$,
respectively, when $g=ia$. Tables \ref{tab:Mathieu_e} and
\ref{tab:Mathieu_o} show the critical parameters for the even and
odd solutions, respectively, to the Mathieu equation
(\ref{eq:Mathieu}). They approximately follow a straight line of
the form $a_{n}=0.582e_{n}+3.66$. Once again we appreciate that
both $e_{n}$ and $a_{n}$ increase asymptotically as $n^{2}$.

\section{Non-hermitian perturbed three-dimensional rigid rotor}

\label{sec:3D_rotor}

An even more interesting example of rigid rotor is provided by
\begin{equation}
\hat{H}=\hat{L}^{2}-g\cos (\theta ),  \label{eq:rotor_3D}
\end{equation}
where $\hat{L}^{2}$ is the square of the dimensionless
quantum-mechanical angular-momentum operator. This Hamiltonian is
invariant under the antiunitary transformation
$\hat{A}=\hat{U}\hat{T}$ discussed above when $g$ is purely
imaginary.

In order to apply the DM we resort to the set of eigenvectors $\left|
l,m\right\rangle $ of $\hat{L}^{2}$ and $\hat{L}_{z}$:
\begin{eqnarray}
\hat{L}^{2}\left| l,m\right\rangle &=&l(l+1)\left| l,m\right\rangle ,
\nonumber \\
\hat{L}_{z}\left| l,m\right\rangle &=&m\left| l,m\right\rangle ,
\end{eqnarray}
where $l=0,1,\ldots $ and $m=0\pm 1,\pm 2,\ldots ,\pm l$ are the angular
momentum and magnetic quantum numbers, respectively. Every eigenvector $%
\left| \psi \right\rangle $ of $\hat{H}$ can be expanded as
\begin{equation}
\left| \psi \right\rangle =\sum_{i=0}^{N}c_{i}\left| M+i,m\right\rangle ,
\end{equation}
where $M=|m|$ and the coefficients satisfy the recurrence relation\cite{F01}
(and references therein)
\begin{eqnarray}
&&A_{i}c_{i-1}+B_{i}c_{i}+A_{i+1}c_{i+1}=0,  \nonumber \\
&&A_{i}=-g\left[ \frac{i(i+2M)}{4(i+M)^{2}-1}\right]
^{1/2},\;B_{i}=(i+M)(i+M+1)-E.  \label{eq:RR_3D_rec_rel}
\end{eqnarray}
There is also a simple recurrence relation for the secular determinants\cite
{F01} but we do not need it here because we can efficiently obtain $E(g)$
from the roots of $c_{N}=0$ for sufficiently large $N$.

We denote $E_{M,n}$, $M,n=0,1,\ldots $ the eigenvalues of $\hat{H}$ so that $%
E_{M^{\prime },n^{\prime }}=E_{M,n}$ when $M+n=M^{\prime }+n^{\prime }$ and $%
g=0$. The eigenvectors $\left| \psi _{m,n}\right\rangle $ with
$m=\pm M$ are degenerate. In the coordinate representation the
basis set of eigenvectors of $\hat{L}^{2}$ and $\hat{L}_{z}$ are
the spherical harmonics $\left\langle \theta ,\phi \right. \left|
l,m\right\rangle =Y_{l}^{m}(\theta ,\phi )$ that satisfy
$\hat{A}\left| l,m\right\rangle =(-1)^{l}\left| l,-m\right\rangle
$. Besides, it follows from the recurrence relation
(\ref{eq:RR_3D_rec_rel}) that $c_{i,M,n}$ is either real or
imaginary when $i$ is even or odd, respectively. Therefore,
$c_{i,M,n}^{*}(-1)^{M+i}=(-1)^{M}c_{i,M,n}$ and $\hat{A}\left|
\psi _{m,n}\right\rangle =(-1)^{M}\left| \psi _{-m,n}\right\rangle
$. We clearly see that in this case $\hat{A}\left| \psi
_{m,n}\right\rangle \neq \lambda \left| \psi _{m,n}\right\rangle $
but the eigenvalue $E_{M,n}$ is real because $\hat{H}\hat{A}\left|
\psi _{m,n}\right\rangle =E_{M,n}\hat{A}\left| \psi
_{m,n}\right\rangle $ in agreement with the more general condition
for real eigenvalues developed in section~\ref{sec:pseudo_sym}.

\begin{figure}[tbp]
~\bigskip\bigskip
\par
\begin{center}
\includegraphics[width=6cm]{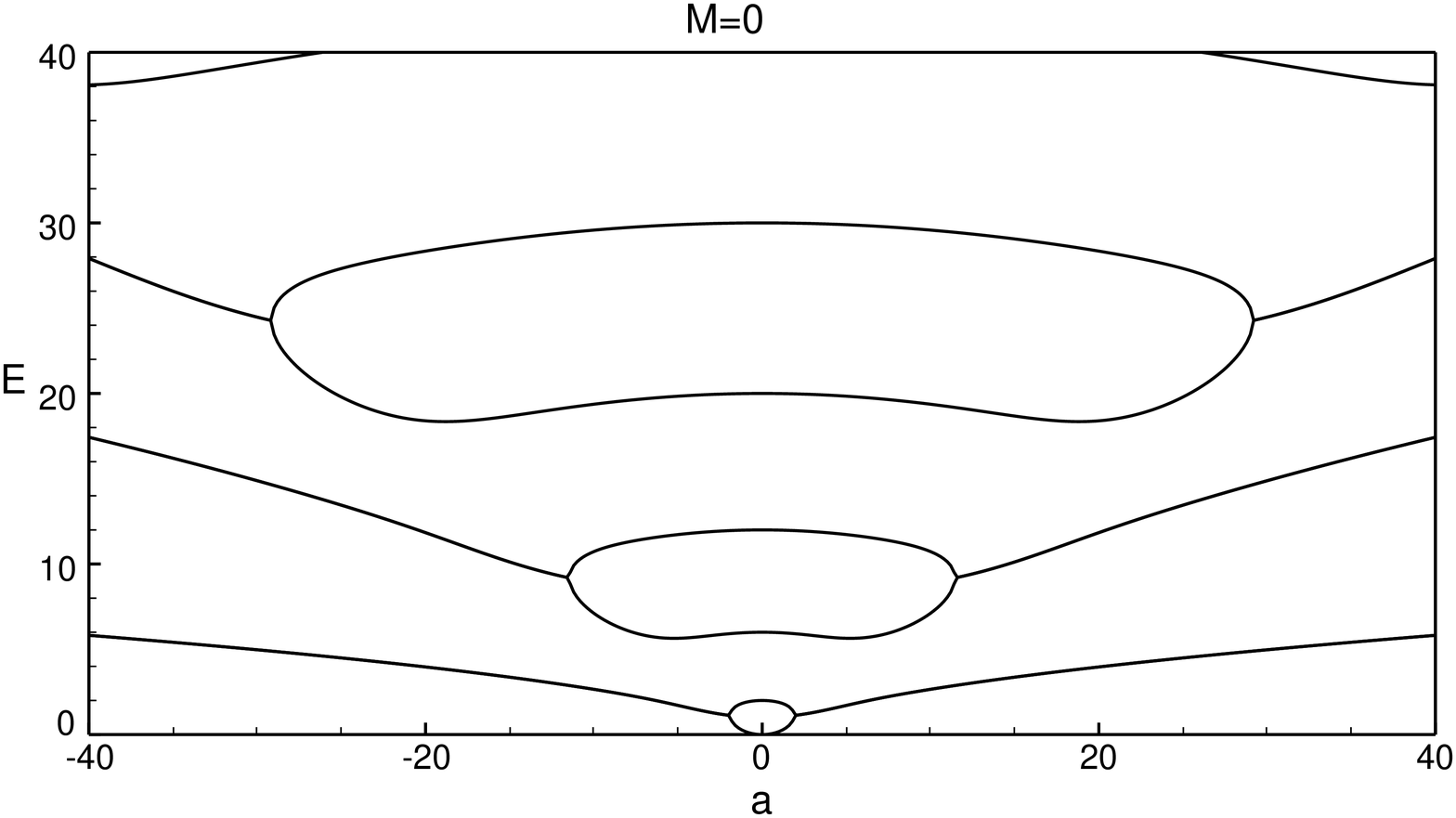} %
\includegraphics[width=6cm]{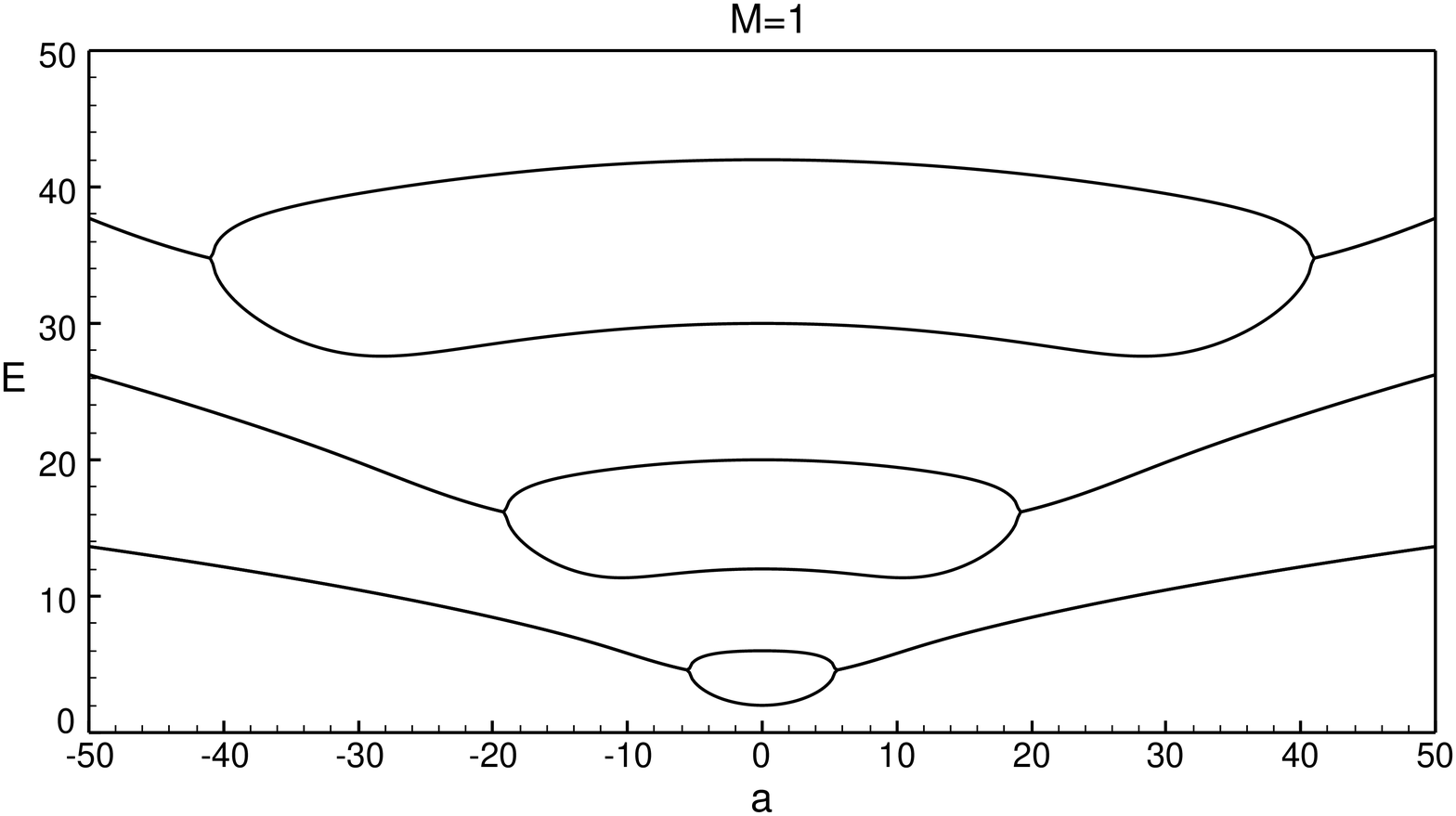} %
\includegraphics[width=6cm]{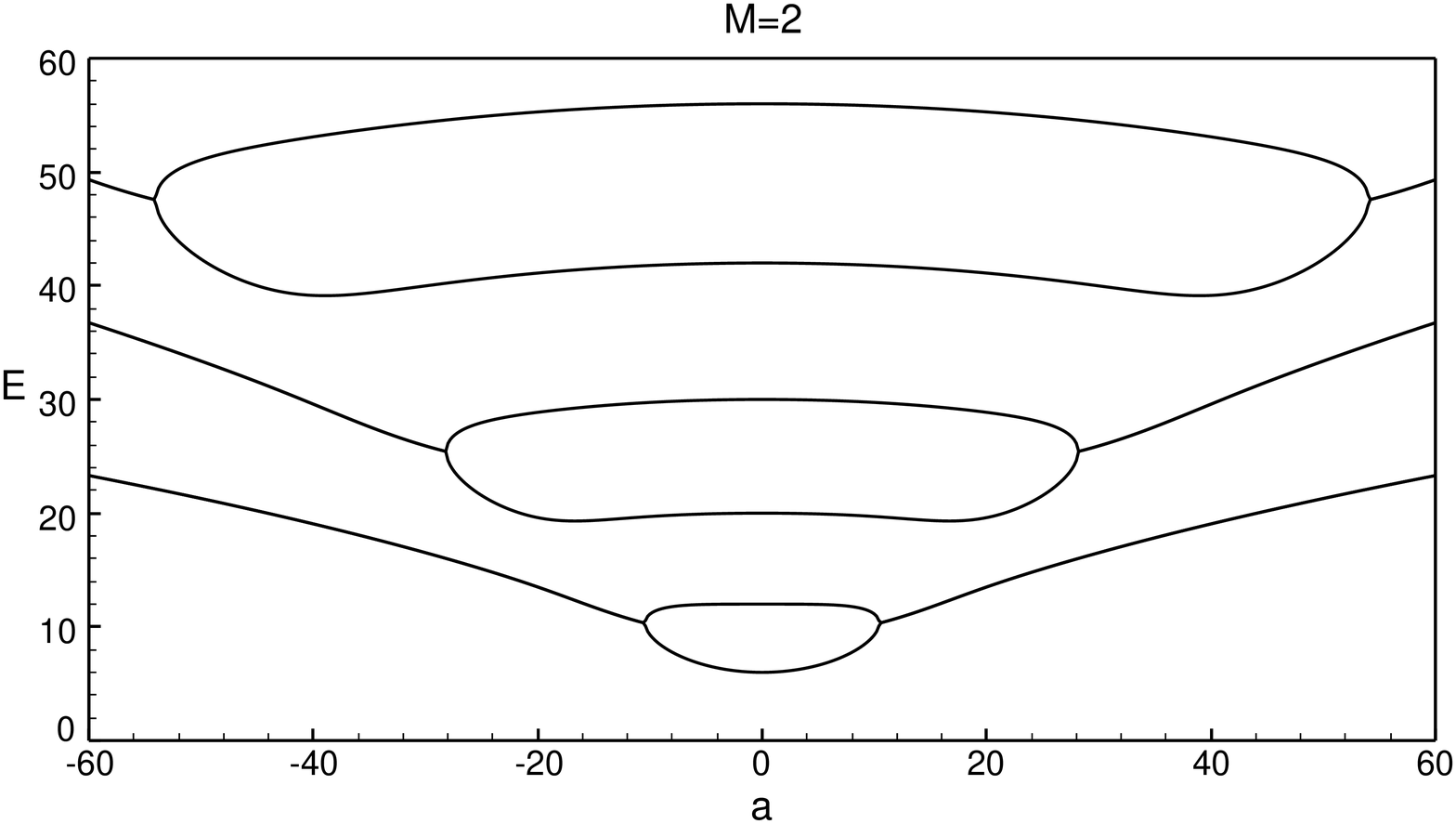} %
\includegraphics[width=6cm]{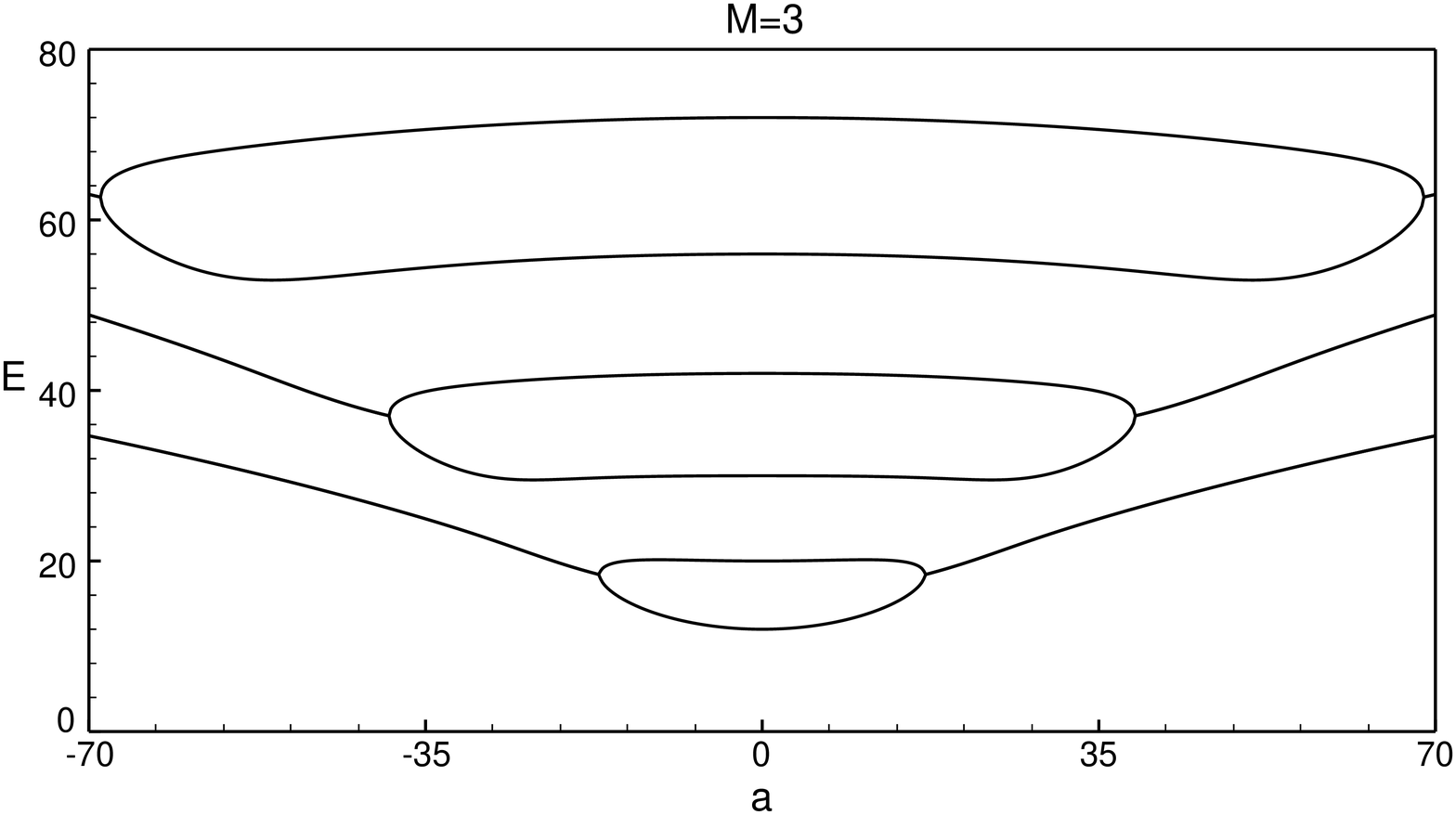} \bigskip
\end{center}
\caption{Eigenvalues $E_{M,n}(ia)$ of the rigid rotor (\ref{eq:rotor_3D}) }
\label{fig:RR3D}
\end{figure}

Fig.~\ref{fig:RR3D} shows the eigenvalues $E_{M,n}$ for
$M=0,1,2,3$ and $n=0,1,2$. It suggests that pairs of eigenvalues
$(E_{M,2n},E_{M,2n+1})$ coalesce at $a=\pm a_{M,n}$ when $g=ia$.
Tables \ref{tab:RR3DM0}, \ref{tab:RR3DM1}, \ref{tab:RR3DM2} and
\ref{tab:RR3DM3} show several critical parameters for $M=0,1,2,3$,
respectively .  In this case we also find a linear relationship
$a_{M,n}=b+ce_{M,n}$ between the critical parameters, where
$c\approx 1.18$, and that they increase asymptotically as $n^{2}$.

\section{Conclusions}

\label{sec:conclusions}

It appears to be clear from the results obtained throughout this
paper that the DM is a remarkably simple and efficient tool for
the calculation of eigenvalues and eigenvectors of a wide class of
PT-symmetric models. In fact, the DM appears to converge more
rapidly than more elaborate approaches\cite {HV13} and seems to be
particularly useful for the calculation of critical parameters and
exceptional points.

The condition for real eigenvalues developed in section~\ref{sec:pseudo_sym}
appears to be more general than the one invoked in earlier studies of the
PT-symmetric Hamiltonians. This fact is plainly illustrated by the perturbed
rigid rotator (\ref{eq:rotor_3D}) for which the commonly used condition for
unbroken symmetry (\ref{eq:unb_sym}) does not hold but the eigenvalues are
real as long as the more general condition (\ref{eq:gen_unb_sym}) applies.

Present numerical investigation suggests that both critical
parameters for the three models (\ref{eq:Hbox}),
(\ref{eq:rotor_2D_eq}) and (\ref {eq:rotor_3D}) behave
asymptotically as $n^{2}$. We may be tempted to conjecture that
this is a general property of such systems but that is not the
case. The analysis of the exactly solvable models with piecewise
constant potentials proposed by Znojil and
collaborators\cite{Z01,ZL01,JZ04} reveals a different behaviour.
In the case of the potential $V(x)=iZx/|x|$, $-1<x<1$, with
Dirichlet or periodic boundary conditions at $x=\pm 1$, the
critical parameters $e_{n}$ and $Z_{n}$ appear to behave
asymptotically as $n^{2}$ and $n$, respectively.

\ack We thank Dr. M. Znojil for useful comments and suggestions that helped
to improve this paper

\begin{table}[tbp]
\caption{Convergence of the DM and OPPQ for $\hat{H}=\hat{p}^2+i\hat{x}^3$}
\label{tab:ix3_conv}
\begin{center}
\par
\begin{tabular}{|D{.}{.}{1}|D{.}{.}{15}|D{.}{.}{15}|D{.}{.}{15}|D{.}{.}{15}|}
\hline \multicolumn{1}{|c}{} & \multicolumn{1}{|c}{DM} & \multicolumn{1}{|c}{OPPQ} &
\multicolumn{1}{|c}{DM} & \multicolumn{1}{|c|}{OPPQ} \\
\hline \multicolumn{1}{|c}{$N$} & \multicolumn{2}{|c}{$E_0$} & \multicolumn{2}{|c|}{$E_1$} \\
\hline

20 & 1.15638348063027 & 1.15720107946295 & 4.10944159217725 & 3.85785039690029 \\
40 & 1.15626708286738 & 1.15626701076546 & 4.10922836311577 & 4.10917078909004 \\
60 & 1.15626707198833 & 1.15626707203003 & 4.10922875272617 & 4.10922884747775 \\
80 & 1.15626707198811 & 1.15626707198786 & 4.10922875280961 & 4.10922875282249 \\
100& 1.15626707198811 & 1.15626707198811 & 4.10922875280965 & 4.10922875280956  \\
\hline \multicolumn{1}{|c}{} & \multicolumn{2}{|c}{$E_2$} & \multicolumn{2}{|c|}{$E_3$} \\
\hline
20 & 7.79277572798155 & 7.27293255888356 & 10.3897589647850 & 10.11399345333521 \\
40 & 7.56228430688572 & 7.56274007348397 & 11.3137218751498 & 11.44673034474738  \\
60 & 7.56227386040027 & 7.56226879749661 & 11.3144217612385 & 11.31447586752061  \\
80 & 7.56227385497881 & 7.56227386127323 & 11.3144218200804 & 11.31442184225783  \\
100& 7.56227385497882 & 7.56227385497590 & 11.3144218201957 & 11.31442182025857  \\
\hline

\end{tabular}
\end{center}
\end{table}

\begin{table}[tbp]
\caption{Critical parameters for the oscillator (\ref{eq:Hcubic})}
\label{tab:CP_cubic}
\begin{center}
\par
\begin{tabular}{|D{.}{.}{1}|D{.}{.}{20}|D{.}{.}{20}|}
\hline \multicolumn{1}{|c}{$n$} & \multicolumn{1}{|c}{$e_n$} & \multicolumn{1}{|c|}{$a_n$} \\
\hline

0& 1.28277353565056613093 & -2.61180935658887732269 \\
1 & 4.18138810077014360384 & -5.37587963413369849339 \\
2 & 7.47676353160394726567 & -7.81513358112177472963 \\
3 & 11.03766256181169489101 & -10.07564704682307238859 \\
4 & 14.80256612165608800708 & -12.21531517192134682450 \\
5 & 18.73495127980953607811 & -14.26484986999511696653 \\
6 & 22.81035758069971715940 & -16.24312145518034341186 \\
7 & 27.01113795189653614640 & -18.16282077195707147474 \\
8 & 31.32389750099022726315 & -20.03302515800852425790 \\
9  & 35.73808590511219720429 & -21.86052509995057604960 \\
10 & 40.24515595690885890435 & -23.65057685885184848462 \\
11 & 44.83802791988671972061 & -25.40736007787328669330 \\
12 & 49.51073146222982552736 & -27.13427141318176849625 \\
13 & 54.25815672832834563937 & -28.83412125570701326472 \\
14 & 59.07587562778958582850 & -30.50927027533271429436 \\
15 & 63.96001004047454636742 & -32.16172704057811933141 \\
16 & 68.9071323790276889544 & -33.7932195834531855763 \\
17 & 73.91418907962735035 & -35.405249007491647104 \\
18 & 78.9784407229043496 & -36.9991304044614914 \\

\hline

\end{tabular}
\end{center}
\end{table}

\begin{table}[tbp]
\caption{Critical parameters for the oscillator (\ref{eq:Hcuartic})}
\label{tab:CP_cuartic}
\begin{center}
\par
\begin{tabular}{|D{.}{.}{2}|D{.}{.}{20}|D{.}{.}{20}|}
\hline \multicolumn{1}{|c}{$n$} & \multicolumn{1}{|c}{$e_n$} & \multicolumn{1}{|c|}{$a_n$} \\
\hline

0& 3.17338956654721488704 & 3.16903614167472725234 \\
1 & 11.32761640743725703756 & 7.62596008108023132512 \\
2 & 21.47216949764589814716 & 12.11537100311929607154 \\
3 & 33.02428793244591467473 & 16.61105709045349074831 \\
4 & 45.70317143857586670043 & 21.10901685823201530899 \\
5 & 59.33696104179223837682 & 25.60805225319570978500 \\
6 & 73.80750220757362981500 & 30.10768065951909870293 \\
7 & 89.0276216454863021248 & 34.6076704707909122127 \\
8 & 104.9298551095159538 & 39.10789674411541992 \\
9  & 121.460151413651 & 43.6082861575648 \\
10 & 138.5740516383 & 48.10879285437 \\

\hline

\end{tabular}
\end{center}
\end{table}

\begin{table}[tbp]
\caption{Critical parameters for the box model (\ref{eq:Hbox})}
\label{tab:CP_box}
\begin{center}
\par
\begin{tabular}{|D{.}{.}{2}|D{.}{.}{13}|D{.}{.}{13}|}
\hline \multicolumn{1}{|c}{$n$} & \multicolumn{1}{|c}{$e_n$} & \multicolumn{1}{|c|}{$a_n$} \\
\hline

1 & 7.1085995967646 & 12.3124556722597 \\
2 & 30.70746876678 & 53.18689607587 \\
3 & 70.9578499846 & 122.902601369 \\
4 & 127.862609648 & 221.464536296 \\
5 & 201.42215453 & 348.87340541 \\
6 & 291.6365885 & 505.1293887 \\
7 & 398.5059476 & 690.2325483 \\
8 & 522.030247 & 904.182911 \\
9 & 662.209493 & 1146.98049 \\
10 & 819.043691 & 1418.625286 \\
11 & 992.532842 & 1719.11731 \\
12 & 1182.67695 & 2048.4566 \\
13 & 1389.476009 & 2406.643039 \\
14 & 1612.93003 & 2793.67675 \\
15 & 1853.0390 & 3209.5577 \\
16 & 2109.80293 & 3654.28585 \\
17 & 2383.22182 & 4127.8613 \\

\hline

\end{tabular}
\end{center}
\end{table}

\begin{table}[tbp]
\caption{Critical parameter $a_1$ from the perturbation series}
\label{tab:PT_Crit}
\begin{center}
\par
\begin{tabular}{|D{.}{.}{2}|D{.}{.}{8}|}
\hline \multicolumn{1}{|c}{$k$} & \multicolumn{1}{|c|}{$a_1(k)$} \\
\hline

9&  12.31814954     \\
19& 12.31354496    \\
29& 12.31290237     \\
39& 12.31269737    \\
49& 12.31260686     \\
59& 12.31255909    \\
69& 12.31253084     \\
79& 12.31251277    \\
89& 12.31250051     \\
99& 12.31249181     \\

\hline

\end{tabular}
\end{center}
\end{table}

\begin{table}[tbp]
\caption{Critical parameters for the even states of the Mathieu equation (%
\ref{eq:Mathieu})}
\label{tab:Mathieu_e}
\begin{center}
\par
\begin{tabular}{|D{.}{.}{29}|D{.}{.}{29}|}
\hline  \multicolumn{1}{|c}{$e_n$} & \multicolumn{1}{|c|}{$a_n$} \\
\hline

2.08869890274969540742210705005   &    1.46876861378514199230729308986   \\
27.3191276740344351613697285995   &    16.4711658922636564062419622945   \\
80.6582642367217733231182880374   &    47.8059657025975746007950854808   \\
162.107021116501331382763597087   &    95.4752727072182593469528060868   \\
271.665574614890515399359662310   &    159.479212669357057187230627715   \\
409.333979844643194402422763806   &    239.817810495650789094138995905   \\
575.112259376089614140747231520   &    336.491073930202402676797136666   \\
769.000424132277697815886582932   &    449.499006061556590915787589874   \\
990.998480035440536142042914292   &    578.841608335703329386650346074   \\
1241.10643057248550485513720070   &    724.518881510280902738995966517   \\
1519.32427792923873387283500008   &    886.530826016874701071963928710   \\
1825.65202354516187327873747825   &    1064.87744211774801292171298635   \\
2160.08966840670531841142295477   &    1259.55872998069003603099135522   \\
2522.63721321244381656497120868   &    1470.57468971764989052881373681   \\
2913.29465847091131070672526973   &    1697.92532140597167537370468979   \\
3332.06200456108886585906171039   &    1941.61062510068817839368674171   \\
3778.93925177117865219161016434   &    2201.63060084195557145589058358   \\
4253.92640032424228639948323784   &    2477.98524865972002792020985438   \\
4757.02345039561656754162219537   &    2770.67456857674245292672357391   \\
5288.23040212502649969662284873   &    3079.69856061061467026098084225   \\

\hline

\end{tabular}
\end{center}
\end{table}

\begin{table}[tbp]
\caption{Critical parameters for the odd states of the Mathieu equation (\ref
{eq:Mathieu})}
\label{tab:Mathieu_o}
\begin{center}
\par
\begin{tabular}{|D{.}{.}{29}|D{.}{.}{29}|}
\hline  \multicolumn{1}{|c}{$e_n$} & \multicolumn{1}{|c|}{$a_n$} \\
\hline

11.1904735991293865896020980123 &   6.92895475876018147964342787950      \\
50.4750161557597516452005364504 &   30.0967728375875542000339071418      \\
117.868924160843684783814608183 &   69.5987932768953947914148570394      \\
213.372568637479279993815862834 &   125.435411314308272709560436718      \\
336.986043950205287207567051913 &   197.606678692480922034682411560      \\
488.709384475887730016940247407 &   286.112608761678078262070275163      \\
668.542605654162967762763559437 &   390.953206295596779988894940683      \\
876.485715432799125784653813063 &   512.128473373035028002129394632      \\
1112.53871831587949363459807537 &   649.638411028231983524090563574      \\
1376.70161704521717624727857705 &   803.483019827838526685397002241      \\
1668.97441338489968248739617901 &   973.662300105893000632623663802      \\
1989.35710852120412013890118346 &   1160.17625207096144146935385017      \\
2337.84970328115567018520806164 &   1363.02487585944750618204515605      \\
2714.45219825889803861462924172 &   1582.20817156403242791167234121      \\
3119.16459389224590002123285708 &   1817.72613924973691186689894477      \\
3551.98689051091222035608251051 &   2069.57877896343363081378199317      \\
4012.91908836792421411986633224 &   2337.76609073971215299835857492      \\
4501.96118766068778990529756079 &   2622.28807460462195641810433326      \\
5019.11318854547174684532365214 &   2923.14473057813385462524614741      \\
5564.37509114759075160843417379 &   3240.33605867580166867491717218      \\
6137.74689556870672683751212500 &  3573.86205890991029411196172754       \\
6739.22860189215699095301722791 &  3923.72273129028543611909139321       \\
7368.82021018690440596500439191 &  4289.91807582487536028648347609       \\

\hline

\end{tabular}
\end{center}
\end{table}

\begin{table}[tbp]
\caption{ Critical parameters $e_{0,n}$ and $a_{0,n}$ for the
rigid rotor (\ref{eq:rotor_3D})} \label{tab:RR3DM0}
\begin{center}
\par
\begin{tabular}{|D{.}{.}{29}|D{.}{.}{29}|}
\hline  \multicolumn{1}{|c}{$e_{0,n}$} & \multicolumn{1}{|c|}{$a_{0,n}$} \\
\hline

1.11850860747789604879129584124& 1.89945169187324547365901350058 \\
9.18271110777602614314692313478& 11.4469373135041414112902268409 \\
24.2743374650550706797661994079& 29.1570364187843312750194317104 \\
46.3934021737006494552531157256& 55.0338230301496191682912997241 \\
75.5399201827232487162615410925& 89.0777654885162317220462901731 \\
111.713900380652159264985603687& 131.288974351497399022703273189 \\
154.915347674365981794080626693& 181.667486575132082638246802870 \\
205.144264889286269484450927321& 240.213317402561182509835277416 \\
262.400653742394747193838743959& 306.926474080239226003553463095 \\
326.684515329643806117627299256& 381.806960425731499957841447183 \\
397.995850380696676312538960541& 464.854778613145237343458945801 \\
476.334659399023202928517978051& 556.069929958740540140414381845 \\
561.700942742735329186201508058& 655.452415299711583081531438128 \\
654.094700673266377544747681781& 763.002235190629698442777914673 \\
753.515933385805509695023538647& 878.719390011577165280746746226 \\
859.964641028959999602185575728& 1002.60388003069997387686020878 \\
973.440823717829191452457184709& 1134.65570544194626000993141955 \\
1093.94448154292133694845436910& 1274.87486638866994863623150054 \\
1221.47561457637370003885489988& 1423.26136297886278858803945222 \\
1356.03422287637955815251177760& 1579.81519529525700149353590512 \\
1497.62030649039627961298586330& 1744.53636340218999765413768412 \\
1646.23386545750814120158073731& 1917.42486735037046169355652365 \\
1801.87489981019236211810598331& 2098.48070718025182357309767219 \\
1964.54340957565684861430612511& 2287.70388292446187324632564614 \\
2134.23939477686596016391844190& 2485.09439460958034550662389867 \\
2310.96285543333590341463022204& 2690.65224225745820774568590806 \\
2494.71379156175787158357548067& 2904.37742588620969831042658859 \\

\hline

\end{tabular}
\end{center}
\end{table}

\begin{table}[tbp]
\caption{ Critical parameters $e_{1,n}$ and $a_{1,n}$ for the
rigid rotor (\ref{eq:rotor_3D})} \label{tab:RR3DM1}
\begin{center}
\par
\begin{tabular}{|D{.}{.}{29}|D{.}{.}{29}|}
\hline  \multicolumn{1}{|c}{$e_{1,n}$} & \multicolumn{1}{|c|}{$a_{1,n}$} \\
\hline

4.55877886725924641484810680290& 5.41369967947421154076411805664 \\
16.1375907539446796948176665321& 19.0366539366410365977084445332 \\
34.7430624620380644472582121023& 40.8287653735067375076502059995 \\
60.3758571013293211953590679065& 70.7886035789264845417420851348 \\
93.0360844044641996314893055308& 108.915912141697487819521883860 \\
132.723772524583231646115964443& 155.210615993212798290721507998 \\
179.438930647130853022234048182& 209.672686433039528417476556225 \\
233.181562239157762078784983517& 272.302110446390569333909532764 \\
293.951668731053476959597684378& 343.098881384146395386662172917 \\
361.749250740754368699280246494& 422.062995536098442607803489884 \\
436.574308535710155676251436297& 509.194450686347541286810153760 \\
518.426842224132039104965417532& 604.493245438036181898256235869 \\
607.306851839808768157712103699& 707.959378871064297125726282783 \\
703.214337381672878552754182737& 819.592850356874342990995371324 \\
806.149298832917264169042646291& 939.393659452681323113486417084 \\
916.111736170419252202468288054& 1067.36180583827770608337708586 \\
1033.10164936940069761597136643& 1203.49728927679401104146044559 \\
1157.11903840568460024469637942& 1347.80010958951204915746935249 \\
1288.16390325671987308020948612& 1500.27026663922414661704954168 \\
1426.23624390197230137796623183& 1660.90776031895581341708604925 \\
1571.33606032299333370614104097& 1829.71259054414926566147073681 \\
1723.46335250333125098536155945& 2006.68475724713587692441388188 \\
1882.61812042837198303912549092& 2191.82426037315622876377594946 \\
2048.80036408515554263447292272& 2385.13109987744749454382949289 \\
2222.01008346219172139310274467& 2586.60527572308025767686172501 \\
2402.24727854928656430083016851& 2796.24678787933020022750968876 \\
2589.51194933738457410429640333& 3014.05563632043725171822918588 \\
2783.80409581842809634738596243& 3240.03182102464926859195853280 \\

\hline

\end{tabular}
\end{center}
\end{table}

\begin{table}[tbp]
\caption{ Critical parameters $e_{2,n}$ and $a_{2,n}$ for the
rigid rotor (\ref{eq:rotor_3D})} \label{tab:RR3DM2}
\begin{center}
\par
\begin{tabular}{|D{.}{.}{29}|D{.}{.}{29}|}
\hline  \multicolumn{1}{|c}{$e_{2,n}$} & \multicolumn{1}{|c|}{$a_{2,n}$} \\
\hline

10.3208166747903646568973037932 & 10.4288550159906556880861532857\\
25.4207623327887544466386812765& 28.1582740390332761901432188657\\
47.5418883505162195939968735032& 54.0402932178718093860340915287\\
76.6891413244321657490533380555& 88.0863457427007731538106071227\\
112.863428447014039279577496571& 130.298593278713537202154514528\\
156.065011915265846305454988719& 180.677682922350525891381170392\\
206.293988039332160271025550914& 239.223862208215672094273506007\\
263.550397974137409898781499507& 305.937241931464725274206268463\\
327.834261238895647696546524794& 380.817877474579788921392992095\\
399.145587817865851310066047948& 463.865798896565719257861484418\\
477.484383113449595517060626653& 555.081023588338983061868169447\\
562.850650174688796531422680455& 654.463562138981667213774784976\\
655.244390776764559156247145443& 762.013421271343462077625746903\\
754.665605975802226624019198276& 877.730605405071181726830864340\\
861.114296408136089048224877344& 1001.61511753266849630211224248\\
974.590462458325714246783348069& 1133.66695973221802998481790522\\
1095.09410435669169203890272295& 1273.88613347875288396645703396\\
1222.62522223754679315385272073& 1422.27263983948977169496533497\\
1357.18381617478886848820532514& 1578.82647959970741927897388563\\
1498.76988620408725276163261494& 1743.54765334592907523155490638\\
1647.38343233693679094718818617& 1916.43616152210473017073444202\\
1803.02445456967522723386079799& 2097.49200446830527099050583774\\
1965.69295288932481713767155384& 2286.71518244784416497102175928\\
2135.38892727740046871358920607& 2484.10569566659182256233911612\\
2312.11237771239876092876125824& 2689.66354428692976622447303147\\
2495.86330417142194483897972208& 2903.38872843796560070724063761\\
2686.64170663122979954448053178& 3125.28124822310130537164659655\\
2884.44758506891063512567586423& 3355.34110372570295723343854640\\

\hline

\end{tabular}
\end{center}
\end{table}

\begin{table}[tbp]
\caption{ Critical parameters $e_{3,n}$ and $a_{3,n}$ for the
rigid rotor (\ref{eq:rotor_3D})} \label{tab:RR3DM3}
\begin{center}
\par
\begin{tabular}{|D{.}{.}{29}|D{.}{.}{29}|}
\hline  \multicolumn{1}{|c}{$e_{3,n}$} & \multicolumn{1}{|c|}{$a_{3,n}$} \\
\hline

18.3932656869754919346793973788& 16.8966533642743226378461806280 \\
37.0261276648638864684719643139& 38.7837061748124744181847563089 \\
62.6669859858232821773304748336& 68.7740586891067799587798103556 \\
95.3306215019736927494865643118& 106.915119263007675368240263857 \\
135.020028981874691707132974862& 153.217194854387105286541377321 \\
181.736151824585196120205263944& 207.683669335464793838124652103 \\
235.479364255103716530989139084& 270.315929035220727532388027826 \\
296.249838077915056335119725000& 341.114629238351091770190392543 \\
364.047660655340315855748376441& 420.080112725581765701401257790 \\
438.872879915491109161811406148& 507.212572987544823448482273980 \\
520.725523743932161740530113278& 602.512125938872178124628573208 \\
609.605609143244375059516152739& 705.978844426605700033807188021 \\
705.513146883842805626496171499& 817.612776073132832143973505231 \\
808.448144007708646910693123011& 937.413953057538062250373844424 \\
918.410605243078958607010397392& 1065.38239774458967328397480722 \\
1035.40053383702430388114116904& 1201.51812605992292426362249599 \\
1159.41793206278032717306053303& 1345.82114958635061001956413623 \\
1290.46280153831843226127329649& 1498.29147690625339213241310920 \\
1428.53514343169852409564937760& 1658.92911448450732063468893814 \\
1573.63495859652721769067726115& 1827.73406726311001705562366049 \\
1725.76224766315409747995008471& 2004.70633907016901625781758106 \\
1884.91701110119625582112832348& 2189.84593290656915762534485743 \\
2051.09924926310976011661273016& 2383.15285115035342904111463156 \\
2224.30896241500183043304906063& 2584.62709570470366379109882661 \\
2404.54615075871058072339672775& 2794.26866810660184529365303827 \\
2591.81081444781812982947448859& 3012.07756960765198814361352503 \\
2786.10295359939116923987354036& 3238.05380123490915423799786847 \\
2987.42256830267472568546233513& 3472.19736383716209562367855469 \\

\hline

\end{tabular}
\end{center}
\end{table}

\end{document}